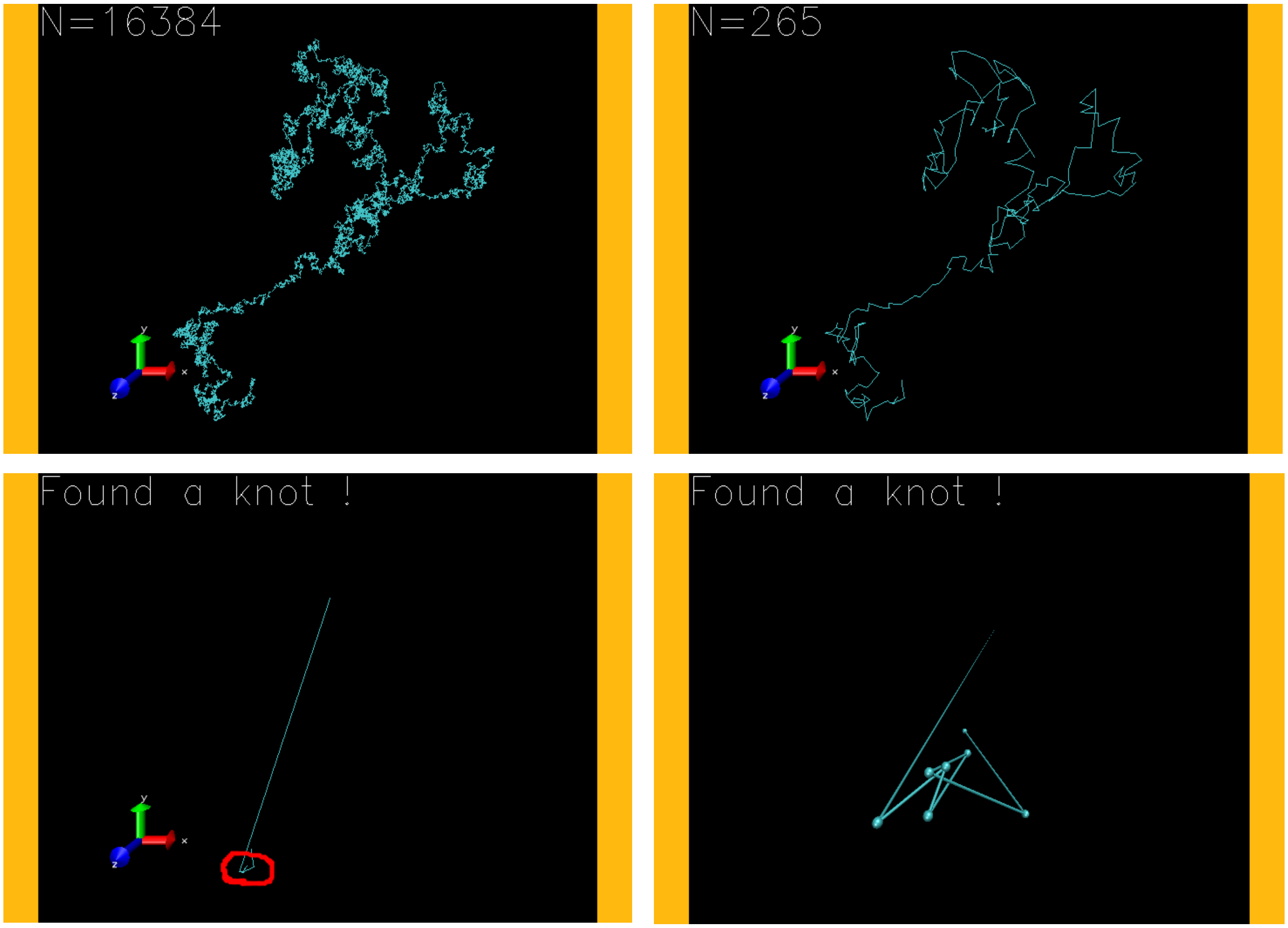

FIG. 1. Knotted bead-spring polymer: Starting configuration with N=16384 beads; after 6 reduction steps (N=265); final configuration after 15 iterations (N=8) with the knotted (trefoil) region circled in red; and magnified.[4] (enhanced online)

# Capturing knots in polymers

Peter Virnau, Mehran Kardar
*Department of Physics, MIT, Cambridge, MA 02139-4307, USA*
Yacov Kantor
*School of Physics and Astronomy, Tel Aviv University,* 69978 Tel Aviv, Israel

Visualizing topological properties is a particularly challenging task. Although algorithms can usually determine if a loop contains a knot, finding its exact location is difficult (and not necessarily well-defined).[1,2]

Here, we apply a reduction method by Koniaris and Muthukumar[3], which was originally proposed to simplify polymers before calculating knot invariants. We start with one end and consider consecutive triangles formed by three adjacent monomers. If the triangle is not crossed by any of the remaining bonds, the particle in the middle is removed. Going back and forth between both ends we proceed until the configuration cannot be reduced any further (see Fig.1).

Although the method is not perfect (sometimes entangled, but unknotted regions remain), it provides us with a valuable impression on the typical number of knots, their respective location and sizes[1].

This work was supported by the DFG grant Vi237/1.

[1] P. Virnau, Y. Kantor, and M. Kardar, J. Am. Chem. Soc., in press (2005).
[2] W. G. Taylor, Nature **406**, 916 (2000).
[3] K. Koniaris and M. Muthukumar, J.Chem.Phys. **95**, 2873 (1991).
[4] Pictures and movie were generated using the VMD visualization package; see W. Humphrey, A. Dalke, and K. Schulten,, J. Molec. Graphics **14**, 33 (1996).

This version also contains a movie of the algorithm.